\documentclass[traditabstract]{aa}  
\usepackage{txfonts}

\usepackage{graphicx,longtable,lscape,natbib,amssymb,amssymb,amsmath}
\bibpunct{(}{)}{;}{a}{}{,} % to follow the A&A style
\newcommand{\ltsima} {$\; \buildrel < \over \sim \;$}  
\newcommand{\gtsima} {$\; \buildrel > \over \sim \;$}  
\newcommand{\lta} {\lower.5ex\hbox{\ltsima}}  
\newcommand{\gta} {\lower.5ex\hbox{\gtsima}}  
\newcommand{\Ha} {H$\alpha$}

\newcommand{\ergs}{\>{\rm erg}\,{\rm s}^{-1}}

\newcommand{\kms}{$\rm{\,km \,s}^{-1}$}

\newcommand{\loiii}{L$_{\rm{\tiny{ [O~III]}}}$}
\newcommand{\lx}{L$_{\rm{\tiny{X}}}$}
\newcommand{\forb}[2]{\mbox{$[{\rm #1\, #2}]$}}
\newcommand{\oiii}{\forb{O}{III}}

\newcommand{\nii}{\forb{N}{II}\,}

\usepackage{color}

\begin{document}

\title{The naked nuclei of LINERs}
\subtitle{} \titlerunning{The naked nuclei of LINERs} 
\authorrunning{B. Balmaverde and A. Capetti}

\author{Barbara Balmaverde\inst{1,2}
\and Alessandro Capetti\inst{3}}

\institute {Dipartimento di Fisica e Astronomia, Universit\`a di Firenze, via
  G. Sansone 1, 50019 Sesto Fiorentino (Firenze), Italy \and INAF -
  Osservatorio Astrofisico di Arcetri, Largo Enrico Fermi 5, I-50125 Firenze,
  Italy \and INAF - Osservatorio Astrofisico di Torino, Via Osservatorio 20,
  I-10025 Pino Torinese, Italy} \offprints{balmaverd@arcetri.inaf.it} 

\date{} 

\abstract{We analyze HST spectra and Chandra observations of a sample of 21
  LINERs, at least 18 of which genuine AGN. We find a correlation between the X-rays and emission lines
  luminosities, extending over three orders of magnitude and with a dispersion
  of 0.36 dex; no differences emerge between LINERs with and without broad
  lines, or between radio-loud and radio-quiet sources. The presence of such a strong correlation is remarkable considering 
that for half of the sample the X-ray luminosity can not be corrected for local absorption.
This connection is
  readily understood since the X-ray light is associated with the same source
  producing the ionizing photons at the origin of the line emission. This
  implies that we have a direct view of the LINERs nuclei in the X-rays: the
  circumnuclear, high column density structure (the torus) is absent in these
  sources. Such a conclusion is also supported by mid-infrared data.  We
  suggest that this is due to the general paucity of gas and dust in their
  nuclear regions that causes also their low rate of accretion and low
  bolometric luminosity.}

\keywords{galaxies: active -- galaxies: nuclei}

\titlerunning{The naked nuclei of LINERs} 
\authorrunning{B. Balmaverde and A. Capetti}
 \maketitle

\section{Introduction}
\label{intro}
The presence of low ionization nuclear emission line regions (LINERs),
initially identified by \citet{heckman80}, is very common in galaxies but
their nature is still matter of debate. This is because several processes,
beside an active nucleus, produce a low ionization spectrum, such as shocks (as
initially proposed by \citealt{heckman80}) or evolved stellar populations
(\citealt{trinchieri91} and \citealt{binette94}).
 In high spatial resolution X-ray images, about 60\% of LINERs 
 show a nuclear compact source (\citealt{gonzalez09}). However the emission from the SMBH
 can account for only $\sim$ 60\% of the ionizing photons and therefore the AGN emission alone is not enough to
  power the optical emission lines (\citealt{flohic06}, \citealt{eracleous10}). These results
 are confirmed by 
the CALIFA
survey: the integral field spectroscopy images show the presence of extended regions characterized by a low ionization
spectrum; however, the radial emission-line surface brightness profiles are
inconsistent with ionization by a central point-source and this is most likely
due to the effects of post-AGB stars \citep{singh13}. It is then unclear
whether LINERs (or which portion of this heterogeneous class) are powered by
accretion onto a supermassive black hole.

Even considering the LINERs that are genuine active galactic nuclei (AGN), we
then still need to understand how they compare to the general AGN
population. For example, whether the orientation based unified model
(e.g. \citealt{antonucci93}) applies also to LINERs, i.e., if the detection of
a broad line region (BLR) is related to the effects of a
circumnuclear obscuring structure \citep{balmaverde14}.

A widely used tool to investigate the nature of AGN is the ratio between the
X-rays and the line luminosities. \citet{maiolino98} found that the $L_{\rm
  X}/L_{\rm[O III]}$ ratio differs by a factor $\sim$100 in Seyferts type 1
and type 2, the former class showing the largest values. This is interpreted
as evidence of nuclear absorption in the X-rays emission in Seyferts 2, since
the \oiii\ emission is a robust (and isotropic) estimator of the AGN
bolometric luminosity. The study by \citet{risaliti99} indicate that in the
Seyferts 2 the local column density, $N_{\rm H,z}$, exceeds $10^{23}$
cm$^{-2}$ in all but two of 35 sources considered. Only when absorption is
taken into account, Seyferts 2 show multiwavelength properties analogous to
those of Seyferts 1. For example, \citet{krabbe01} found a tight correlation
between the 10 $\mu$m (similarly to the emission lines, the MIR light is
radiated almost isotropically) and the X-ray luminosities, followed by both
Seyferts 1 and 2. In most LINERs the quality of the X-ray spectra is
insufficient to measure $N_{\rm H,z}$ directly and they span a very large wide
range of $L_{\rm X}/L_{\rm[O III]}$ values \citep{gonzalez09}, even broader
than in Seyfert galaxies. We still need to establish how LINERs fit into this
scheme.

These issues can be explored in greater depth by studying the nuclear emission
in LINERs at the highest available spatial resolution. Indeed, a key problem
with LINERs is that they are faint objects and thence subject to
strong contamination from the host galaxy. In this Paper we perform an
analysis on a sample of LINERs based on HST and Chandra data. Our aim is to
isolate their genuine nuclear contribution in the X-rays and to perform the
best possible decomposition between the galactic and (if present) the AGN
emission lines.

\section{Sample selection and data analysis}
\label{sample}

\begin{table*}
\caption{Multiwavelength properties of the sample.}
\begin{tabular}{|l|c|c|c|c|c|c|c|c|c|c|r|c|c|}
\hline
Name &  D & Type & & $L_{\rm [O~III]}$ & $L_{\rm [O~III]}$
 & $L_{\rm [N~II]}$
 & $EW_{\rm [O~III]}$ & $L_{\rm X}$& $N_{\rm H}$ {\tiny } &
Ref. & $L_{\rm radio}$& $L_{\rm 12 \mu m}$ \\
& & & &  {\tiny (g.b.)} & {\tiny (HST)} & {\tiny (HST)}& & & (10$^{22}$) & & & \\
\hline
NGC~0315 &   65.8& 2 & RL & 39.44 &    39.63 & 39.82& 130  &    41.8 &  1.06 & a & 37.56  & -- \\ 
NGC~1052 &   17.8& 2 & RL & 40.10 &    39.06 & 39.27& 280  &    41.1 &  12.9 & a & 37.14  & 42.1 \\ 
NGC~1961 &   53.1& 2 & RQ & 39.11 &    --    & 39.08&  29  &    40.4 &  0.8 & b & 34.88  &   -- \\ 
NGC~2787 &   13.0& 2 & RQ & 38.37 &    37.75 & 38.63& 9.8  &    39.1 &  --   & c & 34.33  &   -- \\ 
NGC~3031 &    1.4& 1 & RQ & 37.72 &    --    & 37.55&  28  &    39.4 &  --   & d & 33.76  & 39.9 \\
NGC~3368 &    8.1& 2 & RQ & 37.64 & $<$36.40 & 36.36&$<$0.5& $<$38.3 &  --   & c & 33.07  & $<$40.5 \\
NGC~3998 &   21.6& 1 & RQ & 39.62 &    39.48 & 40.00&  70  &    41.7 &  0.01 & a & 35.68  & 42.0 \\ 
NGC~4036 &   24.6& 2 & RQ & 39.16 &    --    & 38.52&  13  &    39.1 &  2.3  & a & 34.21  &   -- \\ 
NGC~4143 &   17.0& 2 & RQ & 38.81 &    38.49 & 39.23&  25  &    40.0 &  --   & c & 34.23  &   -- \\
NGC~4203 &    9.7& 1 & RQ & 38.53 &    38.28 & 38.76&  38  &    40.1 &  --   & c & 34.21  &   -- \\
NGC~4278 &    9.7& 2 & RL & 38.88 &    37.43 & 37.70&  35  &    38.8 &  2.65 & a & 35.18  & 39.9 \\ 
NGC~4450 &   16.8& 1 & RQ & 38.78 &    38.43 & 38.92&  24  &    40.1 &  --& c & 34.14  &   -- \\
NGC~4477 &   16.8& 2 & RQ & 38.82 &    37.44 & 38.36&  4.2 &    38.6 &  --   & c & 33.70  &   -- \\
NGC~4486 &   16.8& 2 & RL & 39.07 &    38.50 & 39.06&  23  &    40.8 &  --   & a & 37.16  & 41.2 \\ 
NGC~4548 &   16.8& 2 & RQ & 38.11 &    37.10 & 37.00&  2.5 &    38.8 &  --   & c & 33.78  &   -- \\
NGC~4579 &   16.8& 1 & RQ & 39.42 &    39.03 & 38.92& 220  &    41.2 &  0.45 & a & 35.16  & 41.8 \\ 
NGC~4636 &   17.0& 2 & RQ & 38.09 &    --    & 36.81&  13  & $<$38.5 &  --   & e & 33.99  & $<$40.8 \\ 
NGC~4736 &    4.3& 2 & RQ & 37.42 &    --    & 35.65& 0.9  &    38.4 &  0.04 & a & 32.75  & 39.8 \\ 
NGC~5005 &   21.3& 2 & RQ & 39.41 &    --    & 38.99&  31  &    40.0 &  --   & f & 33.95  & 41.0 \\ 
NGC~5077 &   40.6& 2 & RL & 39.52 &    --    & 39.11&  60  &    39.7 &$<$0.1 & g & 36.73  &   -- \\
NGC~6500 &   39.7& 2 & RL & 39.90 &    --    & 38.78&  23  &    39.4 &$<$0.69& h & 36.37  &   -- \\
\hline                              
\end{tabular}
\label{table0}
\medskip

\noindent
\small{Column description: (1) name, (2) distance in Mpc, (3) 1 = BLR, 2 = no
  BLR, (4) flag for radio-loud and radio-quiet sources, (5) logarithm of
  \oiii\ luminosity from ground based observations \citep{ho97},
  (6) through (8) logarithm of \oiii\ and \nii\ luminosities and \oiii\ equivalent
  width (in \AA) from the HST observations (when the \oiii\ is not available
  the EW and L  is derived from the \nii\ flux), (9) through (11): logarithm of the
  X-rays luminosity in the 2-10 keV band  and local column
  density (in cm$^{-2}$) from: (a) \citet{gonzalez09}, (b) this paper, (c)
  \citet{balmaverde13}, (d) \citet{ho01b}, (e) \citet{loewenstein01}, (f)
  \citet{dudik05}, (g) \citet{gultekin12} , (h) \citet{terashima03}, (12)
  radio luminosity $\nu L_\nu$ at 15 GHz from
  \citet{nagar05}, (13) logarithm of the 12 $\mu$m luminosity from
  \citet{asmus14}. All the luminosities are in units of $\ergs$}.
\end{table*}

We consider the 60 galaxies robustly classified as LINERs by \citet{ho97} from
their spectroscopic survey of bright nearby galaxies
\citep{filippenko85,ho95}. We select the objects for which both HST/STIS
spectroscopic data and Chandra observations are available. We find 21 objects,
having discarded 5 sources (namely NGC~4261, NGC~4314, NGC~4594, NGC~4138 and
NGC~4374) due to the poor quality of the HST spectra or because the slit was
not centered on the nucleus. The main properties of these galaxies are listed
in Table \ref{table0}.  Albeit small, this sample provides us with a broad
representation of the LINERs population, in terms of luminosity, host galaxy
type, and the presence of objects with and without broad lines; it also
includes both radio-loud (RL) and radio-quiet (RQ) objects.\footnote{We adopted
  the threshold of radio to X-rays ratio of $10^{-4.5}$ defined by
  \citet{terashima03}.}

We analyze the STIS spectra public in the Hubble Legacy Archive (see Table
\ref{table0}) focusing on those including the \oiii$\lambda$5007 line or, when
not available, the \Ha\ and \nii$\lambda6584$ lines. When possible, we combine
multiple observations to remove cosmic rays and bad pixels. From the
calibrated data we extract the nuclear spectrum from a synthetic aperture of
0\farcs15. The fit to the emission lines is performed with the IDL routine
{\sl mpfit} that employs a $\chi^2$ minimization procedure, modelling each line
with a Gaussian profile. The \nii doublet ratio is fixed to the expected value
of 1:3 \citep{humphrey08}, while the widths of \nii\ and \Ha\ are assumed to
be equal. The continuum emission is reproduced with a first degree
polynomial. For the five type 1 AGN we add a broad \Ha\ component with a
skewed gaussian profile  (see \citealt{balmaverde13} and \citealt{balmaverde14} for the methods of spectra reduction and analysis). In Table \ref{table0} we report the luminosities of
the \oiii\ and \nii\ lines and, for comparison, the \oiii\ luminosity measured
by \citet{ho97}.

\begin{figure*}
\centering{
\includegraphics[scale=0.24,angle=-90]{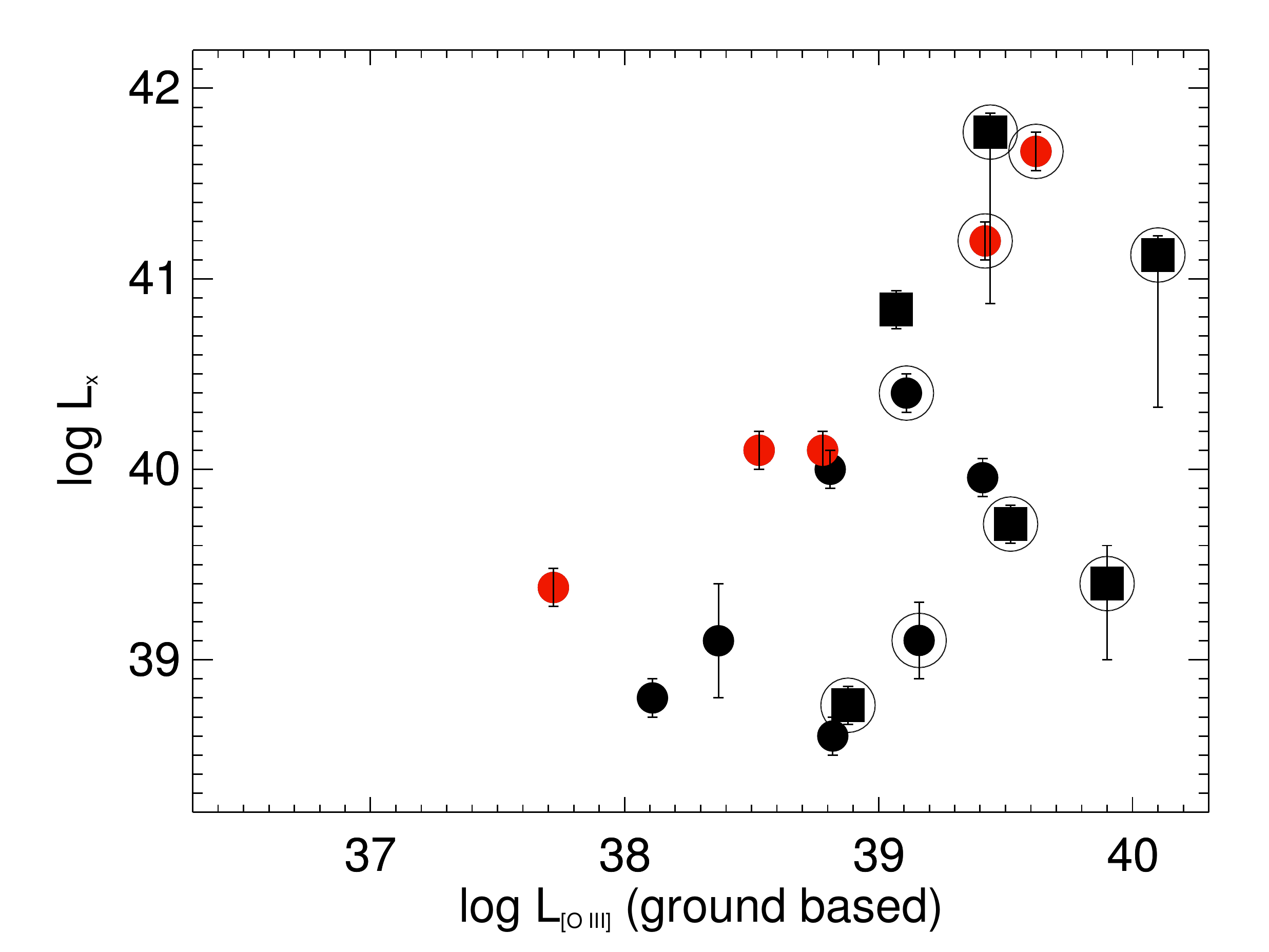}
\includegraphics[scale=0.24,angle=-90]{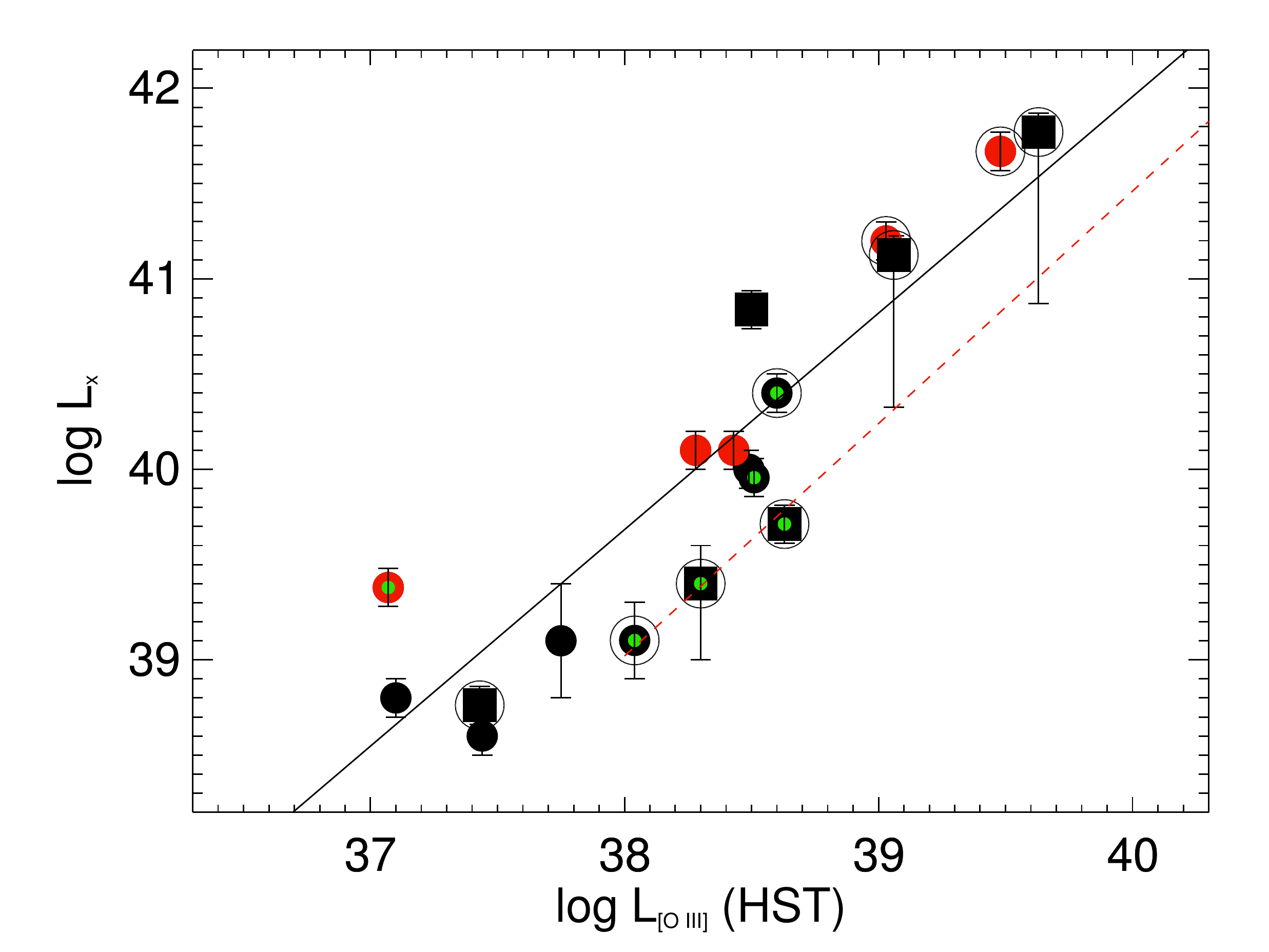}
\includegraphics[scale=0.24,angle=-90]{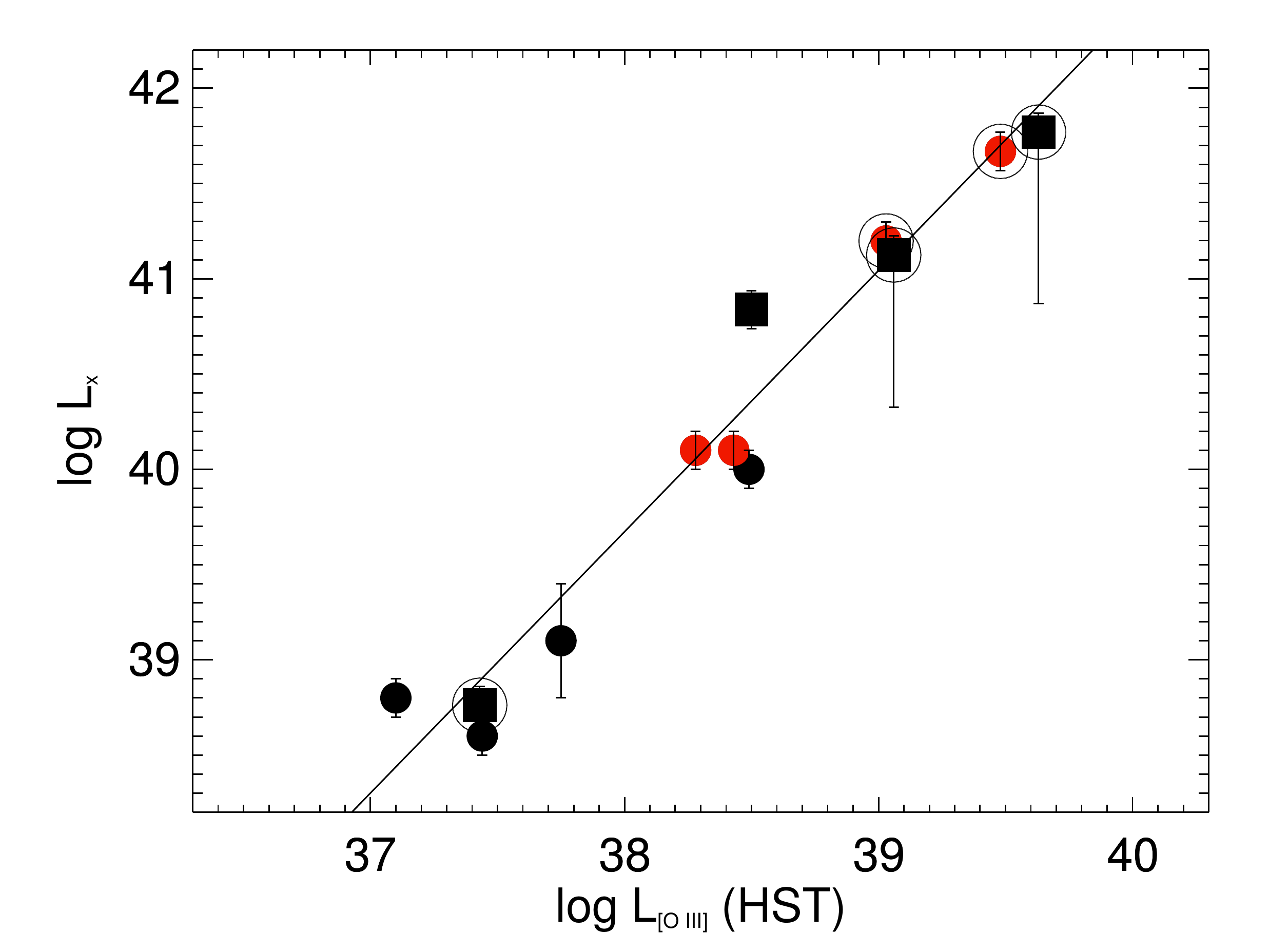}}

\caption{Logarithm of the X-rays luminosity in the 2-10 keV band versus
  \oiii\ luminosity (both in $\ergs$) measured from ground based (left) and
  HST observations (middle), and (right) only for objects with direct
  \oiii\ HST measurements. Type 1 LINERs are the red circles, type 2 objects
  the black ones. The objects for which the \oiii\ luminosity is derived from
  the \nii measurements are marked with the smaller green circles. The squares
  are the RL LINERs. The circles surround objects where $L_{\rm X}$ is corrected for local
  aborption.  In the central and right panel the solid lines represent
  the best linear fit. The dashed red line in the central panel is the
  correlation found for Seyferts by \citet{panessa06}.}
%\caption{Logarithm of the X-rays luminosity in the 2-10 keV band versus
%  \oiii\ luminosity (both in $\ergs$) measured from ground based (left) and
%  HST observations (right). Type 1 LINERs are the red circles, the squares are
%  the RL LINERs. The objects for which the \oiii\ luminosity is derived from
%  \nii measurements are marked with the smaller green circles. The solid line represents the best linear fit. The dashed red
%  line is the correlation found for Seyferts \citep{panessa06}.}
\label{lxlo3}
\end{figure*}

For the Chandra observations, all measurements are collected from the
literature. The only exception is NGC~1961 that we analyze following the same
strategy as in \citet{balmaverde13}. In Table \ref{table0} we list the X-rays
nuclear luminosities and the estimates of the local absorbing column density,
$N_{\rm H,z}$.

\section{X-rays and emission lines luminosities.}
\label{gbcfr}

\citet{balmaverde13} show that low luminosity AGN can be isolated from the
general population of line emission galaxies by considering their HST and
Chandra observations. In particular, AGN show the presence of a nuclear X-ray
source and \oiii\ line with an equivalent width larger than 2\AA\ (smaller EW
values can be produced by stellar processes). 18 of the 21 selected LINERs
fulfill both criteria, the exceptions being NGC~3368, NGC~4636, and NGC~4736.

In NGC~3368 the \oiii\ line is not detected and there is no evidence for a
nuclear X-ray source. NGC~4636 is not detected in the Chandra data \citep{flohic06,loewenstein01,gonzalez09}, but its
large \oiii\ EW (13 \AA) suggests that it might be a genuine AGN.  NGC~4736
is a very peculiar source: it shows a broad \Ha\ line, with a width of only
1570 \kms\ and a luminosity of $2.2 \times 10^{37}$ $\ergs$
\citep{Constantin12}. \citet{izotov07} show that broad lines of this low
width and luminosity are often observed in non active galaxies and are
produced by young (luminous blue variable or O) stars. Indeed,
\citet{eracleous02} argue, based on the analysis of its Chandra observations,
that in this source there is no compelling evidence for the presence of an AGN
and that the X-rays are produced by a dense cluster of young stellar
sources. The low \oiii\ EW (0.9 \AA) further supports this conclusion.

We can now proceed to the analysis of the 18 LINERs that are bona-fide
AGN. The best suited optical line to estimate the AGN power is
\oiii$\lambda$5007 that is less affected, with respect to lines of lower level
of ionization, by the effects of stellar sources, in particular of young
stars. Nonetheless, we prefer to include also objects where only \nii\ data
are available in order to increase the sample size. Spectra covering both the
\oiii\ and \nii lines are available for 12 objects of the sample. The median
ratio between the two lines is 0.33, with a dispersion of a factor 2. We use
this value to estimate the \oiii\ intensity for the objects lacking of a
direct measurement. The uncertainties in the lines luminosities are dominated by the accuracy of the absolute calibration of STIS of the order of $\sim$10\% (\citealt{biretta15}). 

The X-rays luminosities have been corrected for absorption in the nine
  sources where the counts in the Chandra data are sufficient to determine the
  local absorbing column density, $N_{\rm H,z}$. For the remaining LINERs the
  spectra are of insufficient quality to perform a detailed spectral
  analysis. The quoted luminosity are obtained by fixing the power law index
  (to 1.7 or 1.8) and $N_{\rm H,z}$ to zero and, consequently, are not
  corrected for local absorption.

Fig. \ref{lxlo3} compares the X-rays luminosity in the 2-10 keV band and the
\oiii\ luminosity measured from the ground (left panel), from HST (middle
panel), and again from HST data, but including only galaxies with direct
\oiii\ measurements (right). 
Fig. \ref{lxlo3} shows that when using the ground based data we see a trend
of increase of $L_{\rm X}$ with $L_{\rm[O III]}$. However, objects with the
same level of line emission can differ by a factor $\sim$300 in X-rays.  As a
result, the ratio $\log (L_{\rm X} / L_{\rm[O III]})$ spans a broad range,
from -0.5 to 2.3.  Conversely, when using the HST measurements, a clear
correlation emerges, extending over about three orders of magnitude. 
We take into account the X-ray luminosity errors, performing  the linear fit to the data with the IDL routine linfit, obtaining:
$$
\log L_{\rm X} = (40.25 \pm 0.03)+ (1.14 \pm 0.04)\times (\log L_{\rm[O~III]} -38.5)
$$
\noindent
The dispersion around the fit is 0.36 dex.  
Since four out of five of the least luminous sources can not be corrected for local absorption, the slope
of the correlation could be overestimated.  Apparently, there is no distinction
between type 1 LINERs with respect to the type 2 or between RL and RQ LINERs.
The $\log (L_{\rm X} / L_{\rm[O III]})$ ratios span from 1.1 to 2.3 with an
average of 1.7 and a dispersion of 0.4 dex. The \lx\ vs. \loiii\ correlation
for LINERs is almost parallel to that found Seyferts \citep{panessa06}, with an
offset of just 0.5 dex. This is likely due to a different shape of the
spectral energy distributions (SED) of LINERs, as suggested by \citet{ho99}, with a
deficit of ionizing photons at a given X-rays luminosity.

If we consider only the objects with a direct measurement of the \oiii\ line,
we obtain an ever tighter relation (see Fig. \ref{lxlo3}, right panel), with a
dispersion of only 0.2 dex. This indicates that some scatter is due to the
indirect estimates of the \oiii\ luminosity based on the \nii\ data. 

\begin{figure}
\centering{
\includegraphics[scale=0.25,angle=-90]{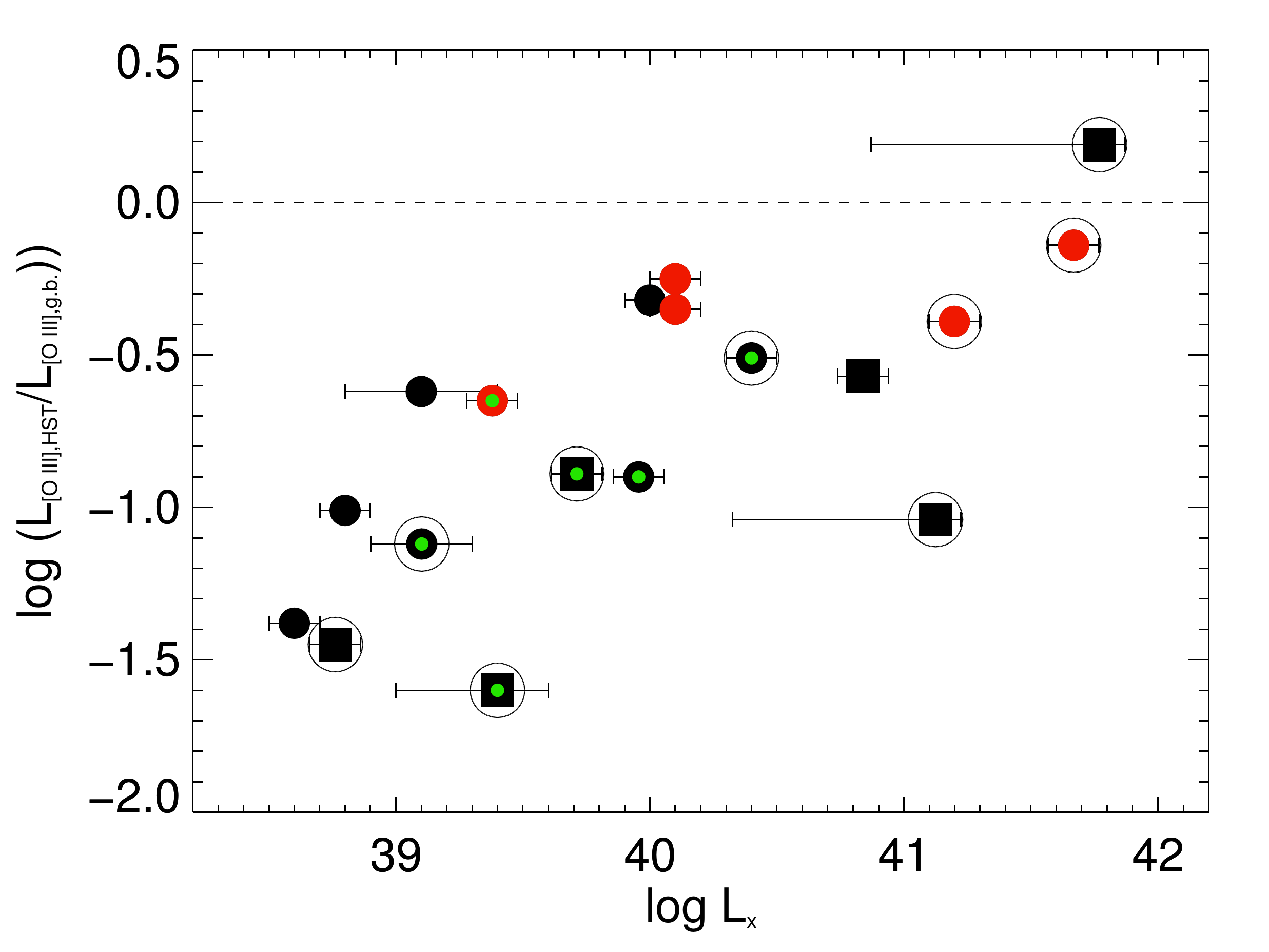}}
\caption{Ratio between the HST and ground based \oiii\ luminosity vs. the
  X-rays luminosity. Symbols as in Fig. 1.}
\label{o3gbhst}
\end{figure}

The difference between the results obtained when using the HST line values as
opposed to the ground based data is due to the fact that, while for the
brightest sources we obtain similar measurements, for the least luminous
LINERs the HST line luminosities are strongly reduced (see
Fig. \ref{o3gbhst}). This is likely due to the fact that the HST data remove
the line emission unrelated to the AGN such as, e.g., that associated with the
stellar processes discussed in the Introduction. This is particularly
important for the least luminous LINERs having the lowest contrast between the
AGN and galactic emission.

\begin{figure}
\centering{
\includegraphics[scale=0.25,angle=-90]{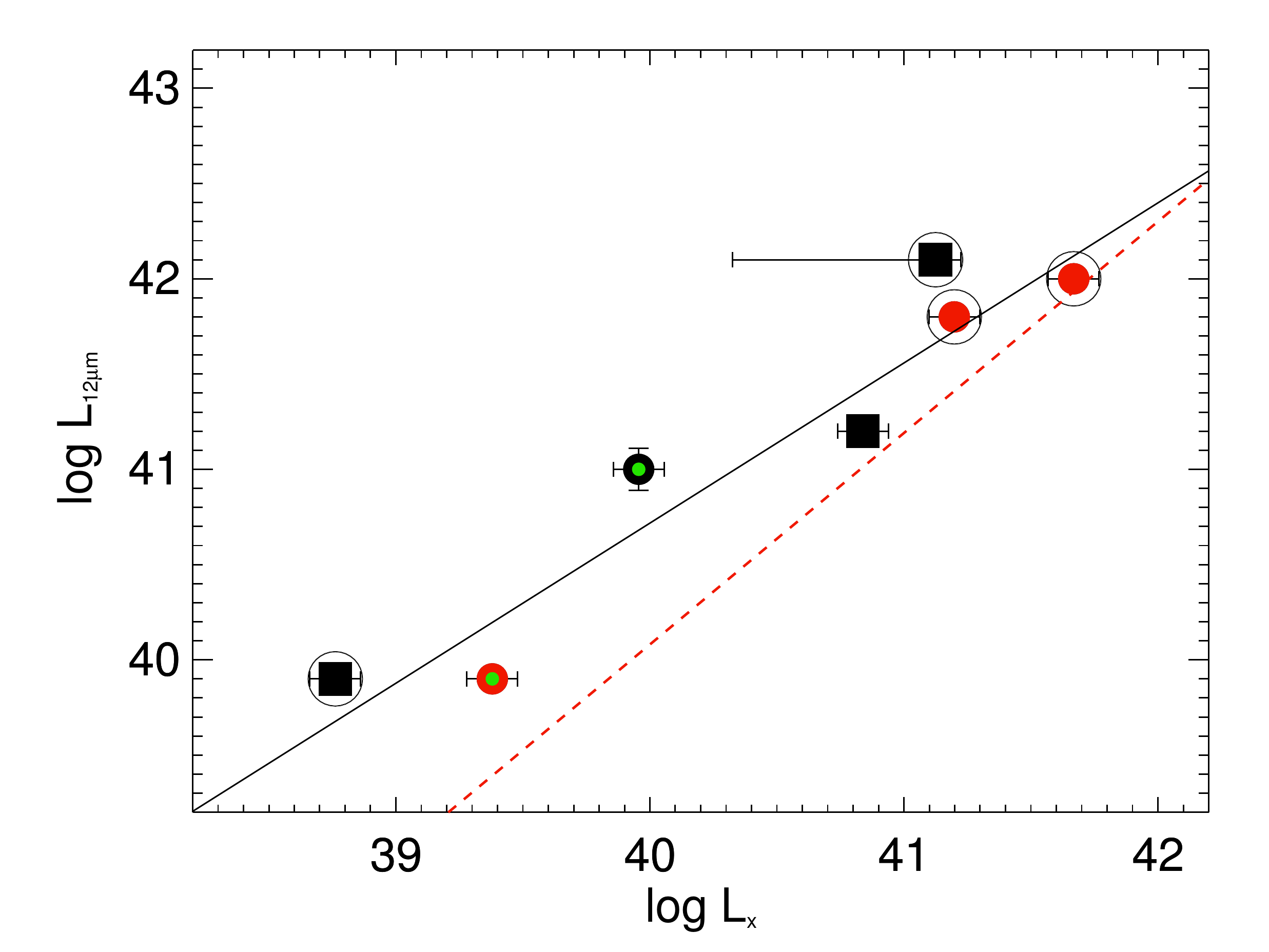}}
\caption{MIR vs. X-rays luminosities for a sub-sample of seven LINERs. The
  solid line is the best linear fit, the dashed red line presents the fit
  obtained for Seyferts by \citet{gandhi09}.}
\label{lxlmir}
\end{figure}

The small spread around the \lx$-$\loiii\ correlation is remarkable
considering the many sources of scatter that are expected to affect it, such
as, for example, the X-rays variability. Most importantly, in half of the
LINERs the X-ray luminosities cannot be corrected for local absorption
  and this, not being included in the model, causes an underestimate of the
  X-rays flux. Simulations performed with PIMMS indicates that the source
flux is reduced by 0.23 dex (the scatter of the \lx$-$\loiii\ correlation) for
$N_{\rm H,z} \sim 10^{22}$ cm$^{-2}$. This can be interpreted as the largest
value of $N_{\rm H,z}$ that can be present is these sources. Note that the
direct estimates available are all (with just one exception) lower than
$N_{\rm H,z} = 10^{22.4}$ cm$^{-2}$.

Another source of scatter is related to the different linear size of the HST
apertures, mainly due to the range of distances among the sources of the
sample. Although 15 galaxies are at similar distances, between 8 and 25 Mpc,
there are much closer and farther objects. This implies that in LINERs the
line measurement is rather stable against different apertures, an effect that
suggests the dominance of a compact emission line region, confined within a
radius smaller that the slit size, typically $\sim$10 pc.

A further support to the idea of a low absorption in LINERs comes from the
comparison of the X-ray and mid-infrared (MIR) luminosities. We collected 12
$\mu$m ground based measurements (to be preferred to the Spitzer data for
their higher spatial resolution) from \citet{asmus14}, available for seven
LINERs. 
In Fig. \ref{lxlmir} we compare the behaviour of our LINERs with respect to the 
relationship found for local Seyferts \citep{gandhi09}. This correlation is
obtained for  a sample of type 1 and 2 Seyferts spanning an X-ray luminosity range (corrected for obscuration) between $\sim$10$^{41}$ and $\sim$10$^{45}$ $\ergs$.
The luminosities in these two bands for LINERs are well correlated and show a
ratio similar to those measured in brighter AGN.
  This again suggests that the X-ray emission is not
significantly absorbed. This result is consistent with the same interpretation
proposed for Seyferts, i.e., that the MIR emission is due to dust heated by
the high energy nuclear photons. The only difference is related to the lower
column density of the circumnuclear gas (and, thence, dust) in LINERs.

\section{Conclusions}
\label{conclusions}

Correlations between emission lines (both considering the broad and narrow
components) and continuum intensity in various bands have been found in many
studies of AGN (e.g., \citealt{adams75,netzer92,mulchaey94}). These are
readily understood since the nuclear light is produced by (or directly
associated with) the accretion disk, that represents the dominant source of
ionizing photons at the origin of the emission lines. Such correlations 
imply that we have a direct view to their nuclei. The AGN where the nucleus is
obscured depart from these correlations. This is the case, for example, of
Seyferts 2 showing, as reported in the Introduction, a much lower ratio
between X-rays and line luminosity with respect to Seyferts 1.

Our results simply recover for LINERs what is already known for
the other classes of AGN. The observed correlation requires that the LINERs
nuclei are ``naked'', i.e., that they are directly visible in the X-rays. This
does not imply the lack of any X-rays absorption, but that the obscuring
material must be optically thin. Indeed, the direct measurements of the local
column density is generally low, reaching only in some sources a few times
10$^{22}$ cm$^{-2}$.
% {\bf We have many LINERs in
%which such measurement is hampered by the Chandra data quality. However we note that, since we do not correct for
%obscuration, if they were heavily obscured
%we would not observe a $L_{\rm X} / L_{\rm[O III]}$ constant ratio}.
%Larger values are excluded also for the many LINERs in
%which such measurement is hampered by the Chandra data quality, as they would
%induce a scatter in the $L_{\rm X} / L_{\rm[O III]}$ larger than observed. 

Of particular interest is the result that the LINERs with and without a BLR
follow the same relation $L_{\rm X} - L_{\rm[O III]}$. Seyferts 2 do not show
a BLR because this is obscured by the torus and these objects show large
X-rays absorption. This is not the case for LINERs and the general lack of a
BLR in these sources (visible only in 5 out 18 objects) cannot be due to
selective obscuration.

Our results contrast with the claim that a large fraction of LINERs is heavily
obscured and even Compton
thick \citep{gonzalez09}, three of such sources belonging to our sample. We
believe that the observed large spread in $L_{\rm X} / L_{\rm[O III]}$, and in
particular the presence of objects with very low ratios on which the claim of
high absorption is based, is due to the heterogeneous measurements of line
emission. These come from both imaging and spectroscopic data and are
generally obtained from very large apertures, when not integrated over the
whole galaxy. The key result of our analysis is that, in these low luminosity
AGN, only the spatial resolution of HST enable us to isolate the genuine AGN
line emission.

Another class of AGN shows ``naked'' nuclei, namely the low luminosity
Fanaroff-Riley type I (FR~I) radio galaxies. For these objects the evidence of
the lack of an obscuring torus is based on the high level of detection of
optical nuclei \citep{chiaberge:ccc} and on the measurements of low $N_{\rm
  H,z}$ in their X-rays spectra \citep{balmaverde06a}, with only four (out of
12) objects showing the presence of significant absorption with values $N_{\rm
  H,z} \sim (0.2 - 6) \times 10^{22}$ cm$^{-2}$. The FR~I are characterized by
emission line ratios typical of gas of low ionization, i.e., they are
radio-loud LINERs. In our sample there are six RL LINERs and they closely
follow the $L_{\rm X}-L_{\rm[O III]}$ relation. This result is intriguing,
since the multiband correlations found for the nuclear emission in FR~I point
to a common non-thermal origin.

LINERs are known to be of lower luminosities with respect to Seyferts
\citep{kewley06}, most likely due to a lower rate of gas accretion. Various
models predict that below a given threshold in Eddington ratio the radiatively
efficient accretion disk typical of powerful AGNs, changes its structure into
an inner hot and radiatively inefficient flow, possibly an advection-dominated
accretion flow, ADAF, (e.g., \citealt{narayan95}) while only at large radii
the standard disk survives \citep{czerny04}.  From an observational point of view, evidence for a truncated disk 
this has been found modelling their SEDs \citep{Nemmen14}.
This might account for their
different  optical lines ratios with respect
to Seyferts \citep{nagao02,ho08}.  In particular the anticorrelation between the hard X-ray photon index and the Eddington ratio in LINERs
is qualitatively consistent with the presence of an ADAF
(\citealt{gu09}, \citealt{younes11}).

A BLR is detected only in a minority of
LINERs and, at least in the brightest sources, this is likely due to the lack
of the high density clouds needed to form this structure. 
For example
\citet{zhang09} found a decrease in intrinsic absorptions of AGNs with decreasing Eddington ratio, suggesting that
the lack of  intrinsic absorption (e.g., a dusty torus or an optically thick disk wind) may be directly related to the lack of a standard accretion disk.
The absence of an
optically thick obscuring structure in LINERs represents a further indication
that these sources are characterized by a general paucity of gas in their
innermost regions with respect to more powerful AGN.

%\bibitem[{{Biretta}{et~al.}(2015)}]{biretta15}
%{Biretta}, J. et~al. 2015, Baltimore:STScI 

\bibliographystyle{aa}  
%\bibliography{my.bib} 

\end{document}